\documentclass[prb, twocolumn, showpacs, numbers, letterpaper]{revtex4-1}
\usepackage{cmap} 
\usepackage{graphicx}
\usepackage{dcolumn}
\usepackage{bm}
\usepackage{natbib,  amsmath, amssymb}
\usepackage{color}
\usepackage{braket}
\usepackage{siunitx}
\usepackage[colorlinks=true, citecolor=blue, linkcolor=blue, urlcolor=blue]{hyperref}
\usepackage[utf8x]{inputenc}
\newcommand{\mtrx}[1]{\boldsymbol{#1}}

\def\@caption@fignum@sep{*}
\newcommand{\colorcaption}[2][]{\caption[ #1]{(color online) #2}}

\begin{document}
\title{Chemical and valence reconstruction at the surface of  SmB$_6$ revealed  with  resonant soft x-ray reflectometry}
\author{V.\,B.\,Zabolotnyy$^1$}\email{volodymyr.zabolotnyy@physik.uni-wuerzburg.de}
\author{K.\,F{\"u}rsich$^{1}$} \altaffiliation[Present address: ]
{Max Planck Institute for Solid State Research
Stuttgart, Heisenbergstraße 1, 70569 Stuttgart, Germany}
\author{R.\,J.\,Green$^{2,3}$}										
\author{P.\,Lutz$^4$}
\author{K.\,Treiber$^4$}
\author{Chul-Hee\,Min$^4$}
\author{A.\,V.\,Dukhnenko$^5$}
\author{N.\,Y.\,Shitsevalova$^5$}
\author{V.\,B.\,Filipov$^5$}
\author{B.\,Y.\,Kang$^6$}
\author{B.\,K.\,Cho$^6$}
\author{R.\,Sutarto$^7$}
\author{Feizhou\,He$^7$}
\author{F.\,Reinert$^4$}
\author{D.\,S.\,Inosov$^8$}
\author{V.\,Hinkov$^1$}

\affiliation{$^1$Experimentelle Physik IV and R\"ontgen Center for Complex Materials Systems (RCCM), Universit\"at W\"urzburg, Am Hubland, 97074 W\"urzburg, Germany}
\affiliation{$^2$Stewart Blusson Quantum Matter Institute, Department of Physics \& Astronomy, University of British Columbia, 2355 East Mall, Vancouver V6T 1Z4, Canada}
\affiliation{$^3$Department of Physics \& Engineering Physics, University of Saskatchewan, Saskatoon, Saskatchewan S7N 5E2, Canada}
\affiliation{$^4$Experimentelle Physik VII and R\"ontgen Center for Complex Materials Systems (RCCM), Universit\"at W\"urzburg, Am Hubland, 97074 W\"urzburg, Germany}
\affiliation{$^5$I. M. Frantsevich Institute for Problems of Materials Science of NAS, 3 Krzhyzhanovsky Street, Kiev 03680, Ukraine }
\affiliation{$^6$ School of Materials Science and Engineering, Gwangju Institute of Science and Technology (GIST), Gwangju, Korea }
\affiliation{$^7$Canadian Light Source Inc., Saskatoon, Canada }
\affiliation{$^8$Institut für Festkörper- und Materialphysik, TU Dresden, D-01069 Dresden, Germany }
\date{\today}

\begin{abstract}
Samarium hexaboride (SmB$_6$), a Kondo insulator with mixed valence, has recently attracted much attention as a possible host for correlated topological surface states.  Here, we use a combination of x-ray absorption and reflectometry techniques, backed up with a theoretical model for the resonant $M_{4,5}$ absorption edge of Sm and photoemission data, to establish  laterally averaged chemical and valence depth profiles at the surface of SmB$_6$.  
We show that upon cleaving, the highly polar (001) surface of SmB$_6$ undergoes substantial chemical and valence reconstruction, resulting in boron termination and a Sm$^{3+}$ dominated sub-surface region. Whereas at room temperature, the reconstruction occurs on a time scale of less than two hours, it takes about 24 hours below 50 K.  The boron termination is eventually established, irrespective of the initial termination.  Our findings reconcile earlier depth resolved photoemission and scanning tunneling spectroscopy studies performed at different temperatures and are important for better control of polarity and, as a consequence, surface states in this system.
\end{abstract}
 
\maketitle

\section{Introduction}

 SmB$_6$ has been extensively studied for more than 50 years. Originally, it had drawn scientific interest as a mixed valence system,\cite{Menth1969, Allen1980, Lawrence1981, Kasuya1994, Wertheim1977182, Guntherodt1982, Travaglini893,  Sampathkumaran1986, Alekseev1993, Chazalviel1976, Beaurepaire1990} later as a Kondo insulator.\cite{Mandrus1994, Cooley1995, Riseborough2000} More recently, when topological insulators started arresting the attention of a substantial part of the solid state community, the material has become famous  again as a potential topological Kondo insulator, in which surface states with unusual properties are to be observed.\cite{Dzero2010, Feng2013, neupane2013surface, jiang2013observation, Xu2014, Kim2014}

Obviously, for any surface state, be it topological or not, the structure and the composition of the surface plays an important role.  For instance,  it was found that depending on the surface termination the surface in-gap states are formed differently and consist mainly of  the Sm 4f states, which indicates the importance of Sm 4f electrons for describing the surface states.\cite{Kim075131} Using ionic liquid gating, it was possible to alter the surface state by modulating its conductivity by as much as 25\%.\cite{Syers2015}  Recalling the valence instability and the ionic character of the Sm--B bond, it was also suggested that the polarity of the surface should have a strong effect on the formation and the type of the surface states.\cite{Sawatzky2013} 

Since even for a Sm compound with an integer bulk valency  a shift of surface valency is not an unusual phenomenon,\cite{Cho15613} it is not surprising that evidence for a modified surface valency in SmB$_6$ and metallic Sm was revealed as early as 1980.\cite{Allen1980} In one of the most recent x-ray absorption studies,  the surface was reported\cite{Phelan2014}  to consist  almost entirely of Sm$^{3+}$. However, the direction in which the Sm oxidation shifts as compared to the average bulk value is not always universal.\cite{Allen1247995} For example, the opposite oxidation shift has been revealed on the surface of SmOs$_{4}$Sb$_{12}$. Similarly, a Sm$^{2+}$-enriched surface was reported for SmB$_6$, though only for  Sm terminated (001) regions.\cite{Jonathan017038}

In this context, a valuable insight is provided by  scanning tunneling microscopy (STM) and spectroscopy (STS), as these methods offer detailed topographical maps\cite{rossler2016surface} for a cleaved surface, and they detect the presence of  reconstructions\cite{Rossler2014} or corrugations,\cite{jiao2016} as well as  in-plane electronic inhomogeneities.\cite{rossler2016surface} However,  surface phenomena, including variations of the average Sm valence and the occurrence of surface states, are not solely bound to the topmost atomic layer.  For example,  a small enhancement in static magnetic fields on the scale of 40--90\,nm was found in a recent muon-spin-relaxation experiment.\cite{Biswas020410}  To study the depth dependent valency on the nanometer scale, one may perform photoemission experiments at different take-off angles, thus modulating the average escape depth and hence the effective probing depth.  Indeed, up to three different layers were reported at the cleaved SmB$_6$ surface in a recent photoemission study.\cite{Lutz2016} However, uncertainties in the universal inelastic mean free path of the photoelectrons caused by such effects as channeling may introduce considerable  errors,\cite{Barrett035427, Lobo245419} so a  more direct method is necessary.

Resonant soft x-ray reflectometry (RXR), combined with x-ray absorption spectroscopy (XAS), is such a method,  which we use here to investigate the cleaved (001) surface of SmB$_6$. This non-destructive approach  allows  us  to assess the chemical and valence profiles of flat samples on the length scale from a fraction to few hundreds of nanometers.\cite{ZHOU1995223, DIETRICH19951} 
 
The paper is organized as follows: in section \ref{Methods}, we describe the sample preparation and experimental conditions, under which the  reflectometry, absorption and photoemission data were collected. In section \ref{stability}, it is shown that the \textit{in situ} prepared (100) sample surface is clean and remains stable during the data collection, without detectable absorption or desorption of residual gases, or any gradual  loss of Sm or B atoms. In section \ref{SVD}, we examine x-ray absorption spectra collected in the total electron yield (TEY) mode, which are commonly used to approximate  the energy dependent x-ray absorption cross section, $\sigma(E)$. We also show that within 0.3\,\% accuracy there are only two types of Sm ions contributing to absorption spectra. In section \ref{CFT} we introduce a crystal-field model,  with which we fit our absorption data. Later in section \ref{profiling} the model is refined and used  to establish chemical density profiles for all atomic species with their respective valencies, including separate profiles for Sm$^{2+}$ and Sm$^{3+}$. Finally the results are discussed and compared to low temperature photoemission in section \ref{discussion} and summarized in section \ref{summary}.
  
\section{Method and  sample preparation}
\label{Methods}

\begin{figure}
\includegraphics[width = \columnwidth]{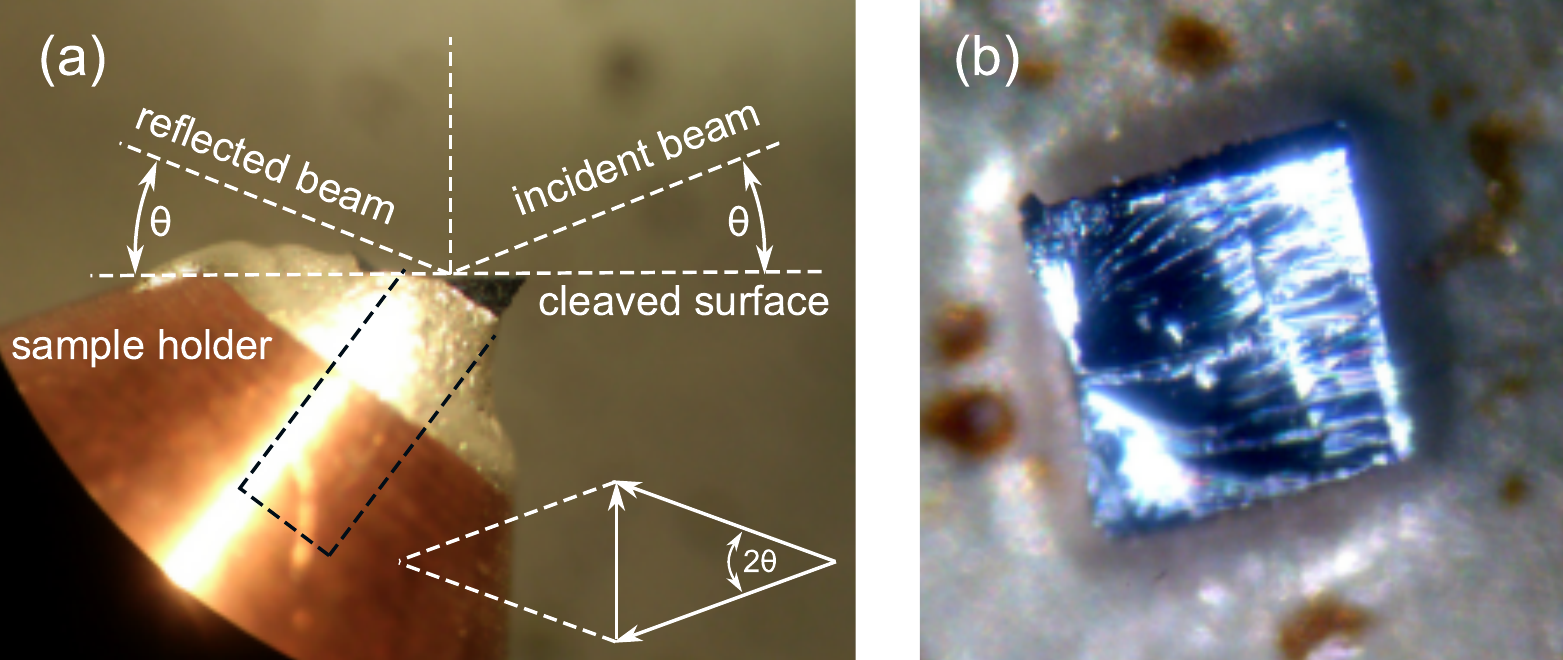}
\colorcaption{(a) Experimental geometry. (b) Optical microscopy of the cleaved sample surface. The sample cross section is 1$\times$1 mm$^2$ }
\label{holders}
\end{figure} 
 
 X-ray absorption and reflectivity measurements were performed using the four-circle UHV diffractometer at the REIXS 10ID-2 beamline of the Canadian Light Source in Saskatoon.\cite{Hawthorn3607438}
 
SmB$_6$ single crystals  were prepared by the floating zone method as described elsewhere.\cite{Friemel2012} The crystals were pre-oriented by Laue diffraction and cut in the form of long rods, which were then glued into conical holders (Fig.\,\ref{holders}) and cleaved at room temperature in the fast-entry chamber  at a pressure of $10^{-8}$ mbar. Immediately  thereafter,  without breaking the vacuum, the samples were  transferred into the measuring chamber of the diffractometer operated under UHV conditions, with a base pressure of better than $2\cdot 10^{-10}$ mbar. After two hours of alignment the actual data were collected, starting with the sample stability and aging tests.

It was found that irrespective of  the pre-orientation and the appropriate cutting, our samples cleaved such as to expose the (001) plane,  which was confirmed by observation of the (001) Bragg peak in the reflectivity spectra. The most comprehensive and complete data set collected on  one of  these  cleaves  is presented and analyzed in the current work.

Since synchrotron-based methods, such as photoelectron spectroscopy or x-ray reflectivity, are sensitive to comparatively large  features reaching in size up to the beam cross section ($\sim 100 \times 100$\,$\mu$m), in Fig.\,\ref{holders}b we show a typical micro-photograph of a cleaved sample.  Although the surface is not ideally flat, it consists of extended flat terraces, thus making specular reflectivity possible.

The soft x-ray photoemission data were collected on single crystals grown in Al-flux.\cite{Regan035159}  The Sm 4f spectra were measured in the fixed mode of a VG Scienta R8000 electron analyzer at the MERLIN Beam line 4.0.3 at the Advanced Light Source (Berkeley, USA) with the photon energy of 70 eV. The B 1s spectra were measured also in the fixed mode of a VG Scienta EW4000 analyzer at the I09 beamline of the Diamond Light Source (Didcot, United Kingdom) with the photon energy of 300 eV. The energy resolutions at both photon energy were better than 100 meV.

\section{Sample stability  and aging }
\label{stability}
Sample aging is an especially notorious issue in photo\-emission,\cite{JongThesis} where just a fraction of a mono-layer of residual gases adsorbed at the surface may critically change the low-energy spectrum by shifting the Fermi level via chemical doping, and/or by wiping out the $k$-resolution  due to the disrupted in-plane long-range atomic order.\cite{Zabolotnyy2012}  In addition, a chemical shift may affect the core-lever spectra.\cite{Lutz2016}

X-ray reflectivity is typically not hindered by a few-nanometer-thick capping layer deposited either intentionally or as an unintended contamination can be taken into account when modeling the RXR data. Nonetheless, a progressively growing or shrinking surface contamination layer would result in a drifting  x-ray  interference pattern, which in turn would render a set of measured spectra mutually inconsistent and impede  the  subsequent extraction of chemical profiles. This is especially true for  samples introduced into the measuring chamber from ambient air due to desorption of CO or other volatile organic products from the solid surface.\cite{Mercier3617, Hoffman2006, Chen20063544, Loubriel1983} Also samples, where the beam changes the stoichiometry at the surface,\cite{Knotek964, Dudy201600046} for example via light-induced oxygen vacancies,\cite{Walker201501556} may pose a problem in this regard.  The recently reported thermal desorption of Yb in YbB$_{12}$ is just another specific example for a compositional surface instability.\cite{Hagiwara2016}
The aforementioned problems may partly be seen not only in reflectivity but also in absorption spectra,  which  undergo a strong  drift on the time scale of several hours. The prevention of these problems was one of the main incentives behind resorting to a more involved \textit{in situ} surface preparation.

\begin{figure}
\includegraphics[width = \columnwidth]{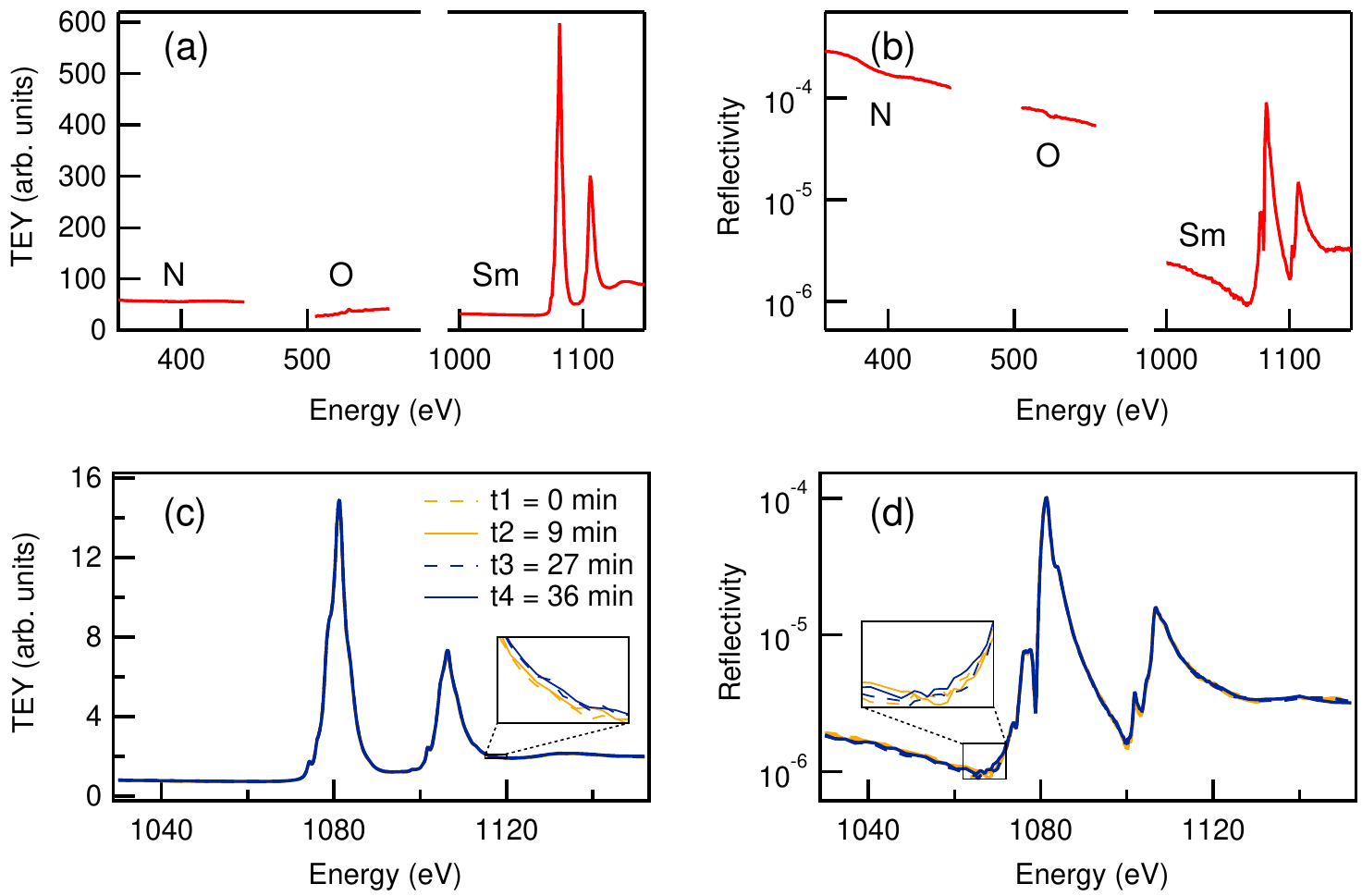}
\colorcaption{ (a,b) TEY  (absorption)  and  reflectivity spectra  over the nitrogen, oxygen and samarium edges. (c,d) Stability of the spectra in TEY and reflectivity modes. The four curves overlap nearly perfectly so that they can hardly be distinguished when overlayed. }
\label{aging}
\end{figure}

To check whether the  \textit{in situ} cleavage solves the problem,  in Fig.\,\ref{aging}(a,b)  we compare the TEY  and RXR spectra  measured over the absorption edges of the dominant residual gases in our fast-entry chamber to the $M_{4,5}$ absorption edge of Sm used as a scale. As one can see, the spectra at the oxygen and nitrogen $K$-edges are practically flat, signifying  virtually no surface contamination. 

Though the data presented in Fig.\,\ref{aging}(a,b) exclude any substantial surface contamination that may start gradually desorbing under the x-ray beam,  there still remains a risk that the beam affects Sm and B compositions at the surface. To test if this is of any relevance for the \textit{in situ} cleaved SmB$_6$, we measured one and the same spectrum several times in a row: first, two sequential spectra collected with nominal beam intensity; then a so-called ``burning'' spectrum measured with about $ 20 \times$ higher beam intensity; and again two spectra with nominal intensity.  The four nominal spectra are shown in Fig. \ref{aging} (c,d) Again, up to small random variations due to statistical noise, all the spectra are found to be identical. Thus we conclude that neither desorption of surface contaminants nor drift in stoichiometry are an issue, and the sample remains stable over the time of data collection.

\section{Decomposing {Samarium} TEY spectra}
\label{SVD}

\begin{figure*}
\includegraphics[width = \textwidth]{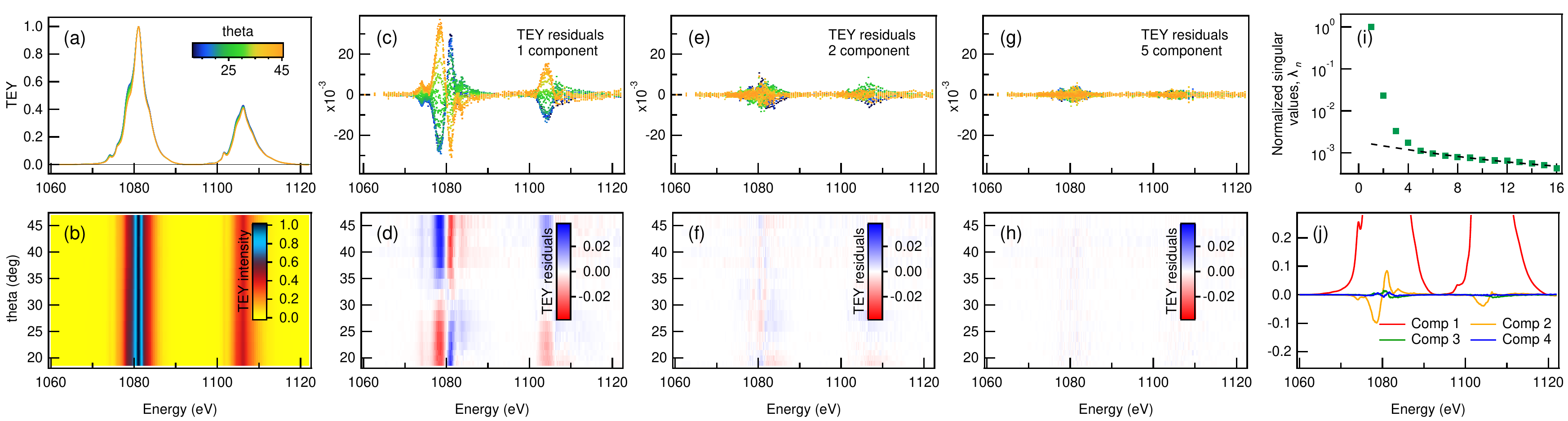}
\caption{SVD decomposition of the TEY data. (a,b) TEY spectra after subtraction of Shirley-like background\cite{Castle1170} and normalization of the peak intensity to one. The false-color plots duplicate the data presented in the line plots and elucidate the  trends in the $\theta$--energy parameter space in a more convenient way. Panels (c,d), (e,f), and (g,h)  show residuals  for one-, two- and five-component representations. (i) Normalized  singular values in decreasing order. (j) The first 6 most significant components.  }
\label{SVD_figure}
\end{figure*}

SmB$_6$ is a mixed-valency system,\cite{Menth1969, Allen1980, Lawrence1981, Kasuya1994, Wertheim1977182, Guntherodt1982, Travaglini893,  Sampathkumaran1986, Alekseev1993, Chazalviel1976, Beaurepaire1990} therefore  generally we expect an absorption spectrum to be the sum of two different components\cite{Min201546}  corresponding to Sm$^{2+}$ and Sm$^{3+}$ ions, respectively. However, in the case of a substantial inhomogeneity the number of different components may be larger.  For example,  at a predominantly Sm terminated surface,  there could be a considerable number of Sm atoms with changed coordination number as compared to the bulk, hence each individual TEY spectrum
would  become a weighted sum of three, or maybe more, components: two corresponding to  the bulk Sm$^{2+}$ and Sm$^{3+}$ while others representing the Sm ions with disrupted coordination or oxidation state:
\begin{equation}
S(\omega) 
= 
\alpha_{1}  C^{1}(\omega) +
\alpha_{2}  C^{2}(\omega) +
\alpha_{3}  C^{3}(\omega) +...
\end{equation} 
The intuition here is that a change in the incidence angle and/or the polarization of the x-rays would affect  the weighting coefficients $\alpha_i$ so that upon recording a set of spectra $\{ S_n(\omega)\}$  with varying contributions of the components $ C_i(\omega)$, we can eventually  find the components themselves, or at least their number, that is the dimensionality  of the basis in which each of the measured spectra can be expanded.  Obviously, the basis formed by the component spectra $\{ C_i(\omega) \}$  is not unique.  However, applying singular value decomposition (SVD), one can objectively  determine the minimal number of \textit{significant} components  
that are relevant at a given quality of the input data.  SVD is a very powerful method used in the analysis of magnetic resonance imaging\cite{Mitra1910370407} and other biological data,\cite{Landahl2013} NEXAFS microscopy and resonant scattering.\cite{Ade2008}  

Suppose that we  have $N$ absorption spectra with $M>N$ points, each measured on the same energy grid $\{ \omega_m \}$,    $m =1,..., M$,  under progressively changing conditions.  The measured data can be placed in a  $M \times N$ matrix, such  that each column contains a particular spectrum. We can always perform SVD decomposition of this matrix, in which we represent $\mtrx{S}$ as 
\begin{equation}
 \mtrx{S}_{M \times N} = \mtrx{U}_{M \times M} \mtrx{W}_{M \times N}\mtrx{V}_{N \times N}^\text{T},
\end{equation} 
were $\mtrx{U}$ and $\mtrx{V}$ are orthogonal matrices and $\mtrx{W} = (w_{i, j}) = (\lambda_i \delta_{i, j})$ is the diagonal matrix, containing the singular values, $\lambda_i$, in decreasing order.
 If the measured spectra are linear combinations of only $p$ components, then  only the first $p$ singular values will be non-vanishing, i.e. all spectra can be expanded in the first $p$ columns of the matrix  $\mtrx{U}$. Since real spectra  also contain statistical noise, the value of the remaining singular values will be determined by the noise level, and for a purely  random matrix the singular values $\lambda_k$ will follow the so-called Marchenko--Pastur distribution,\cite{MR0208649} exhibiting a close-to-exponential decrease with $k$.  Though we cannot exclude the presence of extremely weak components swamped by  the noise,  the approach allows us to draw an objective demarcation line between those components whose singular values  are above the noise bed and those that can be discarded as unjustified by the current signal-to-noise ratio.\cite{Mees1987}
 
It is also worth mentioning that regardless of the actual reasons, considering only the first $p$ singular values provides the best mean-squares-$p$-component fit to the original data, which is the essence of the Eckart--Young theorem.\cite{Eckart1936}
 
In Fig.\,\ref{SVD_figure} we present  the results of an application of  the SVD analysis to a set of TEY spectra measured with $\sigma$-polarized light for  $\theta$ ranging from 18$^\circ$ to 45$^\circ$. As can be seen in Fig.\,\ref{SVD_figure}(c,d), there remains a clear structure  in the residuals with the errors amounting up to 3\% ---therefore a single component is clearly not enough to describe the data. Increasing the number of components to two (panels e and f) results in an  almost  perfect description of the experimental data, with very little residual structure  discernible beneath the noise.  This means: (i) the experimental data can essentially be expanded in two basis spectra and  (ii) the overall variation between individual spectra relative to the  averaged one is  unfavorably small, of the order of $\lambda_2/\lambda_1$ = 0.023, though still much larger than the noise level, which is  given here by $\lambda_5/\lambda_1 \sim10^{-3}$.

From the plot of the singular values [Fig.\,\ref{SVD_figure}(i)]  we can actually discern four different components that still  stand out above the noise bed marked by the dashed line. Recalling that TEY spectra are often distorted by self-absorption effects,\cite{Pompa1997, Nakajima1999,  Achkar2011, Kurian12} one may try to explain  the two `unwanted' components out by arguing that the self-absorption would break down the linearity assumptions\cite{Henneken2000, Gota2000} on which our SVD analysis is based. This is indeed a valid argument, however non-linearity would produce deviations that monotonically change as the total intensity grows with $\theta$, so we would observe largely positive residuals on one side of some optimal $\theta_0$ and largely negative residuals on the other side.  Since this is not the case (see  panel f), there indeed must be at least 4 different components in the data set, presumably due to  slightly different effective crystal fields experienced by  Sm$^{2+}$ and Sm$^{3+}$  at the surface and bulk,  respectively.  Since components number 3 and 4 are very weak as compared to components number 1 and 2 (with ratios $\lambda_3/\lambda_1\sim 3\cdot10^{-3}$, $\lambda_4/\lambda_1\sim 2\cdot10^{-3}$) and to the noise, we can safely disregard them. In Fig.\,\ref{SVD_figure}(j), we plot the first 6 components $C^k(\omega_n) = \lambda_k v_{n,k}$, $k=1,...,6$.

Thus  from the SVD decomposition we can justly conclude that only two components are major contributors to the measured spectra and that the relative gross error, due to discarding other components, as well as due to self-absorption effects is of the order of 0.3\%. We should also emphasize that $C^1(\omega_n)$, $C^2(\omega_n)$ are not the Sm$^{2+}$ and Sm$^{3+}$ spectra themselves,  but some linear combinations of these. 

To find the Sm$^{2+}$ and Sm$^{3+}$ spectra one may  try to include some additional restrictions/assumptions, like non-negativity,\cite{Landahl2013} known branching ratio between the spin--orbit split multiplets,\cite{Thole1998, Borrero2016} etc.  In this particular case, we will, however, rely on the crystal field theory (CFT) calculation as a source of restrictions, whereas the physical parameters, like Coulomb repulsion and transition life-times, will be treated as unknown fit parameters to be obtained from experiment.    

\section{$M_{4,5}$-edge, fit to crystal field calculations }
\label{CFT}
\begin{table*}[t]
\caption{\label{tab:table1} Optimized CFT parameters for Sm$^{2+}$ and Sm$^{3+}$ ions.}
\begin{tabular}{l  r c S[table-format=2.4]S[table-format=2.4]S[table-format=2.4]S[table-format=2.4]S[table-format=2.4]S[table-format=2.4]S[table-format=2.4]S[table-format=2.4]S[table-format=2.4]S[table-format=2.4]S[table-format=2.4]S[table-format=2.4]S[table-format=2.4]}
\hline\hline
\rule{0pt}{3.5ex} \rule[-1.5ex]{0pt}{0pt}
{ion} &{ \qquad state  } & {\quad configuration} &{$F^{(2)}_{ff}$} & {$F^{(4)}_{ff}$} & {$F^{(6)}_{ff}$} & {$\zeta_{4f}$} & {$F^{(2)}_{df}$} & {$F^{(4)}_{df}$} & {$G^{(1)}_{df}$} & {$G^{(3)}_{df}$}& {$G^{(5)}_{df}$} & {$\zeta_{3d}$} \\
\hline \rule{0pt}{2.5ex}
{Sm$^{2+}$} & {initial} & {3d$^{10}$4f$^6$ } & 10.828 & 6.751 & 4.845 & 0.136 &          &          &          &          &        &          \\
{                 } & {final}&{3d$^{9\phantom{1}}$4f$^7$ } & 11.548 & 7.218 & 5.185 & 0.165&   6.701 &  3.075 &  4.670 &  2.734 &   1.888 &   10.514 \\
\hline \rule{0pt}{2.5ex}
{Sm$^{3+}$} & {initial} & {3d$^{10}$4f$^5$ } & 10.950 & 6.873 & 4.945 & 0.152 &          &          &          &          &          &          \\
{                   } & {final} & {3d$^{9\phantom{1}}$4f$^6$ } & 11.548 & 7.260 & 5.227 & 0.180&  7.211 &  3.337 &5.086 & 2.979 &  2.058 &   10.510 \\
\hline\hline
\end{tabular}
\label{params}
\end{table*}

In this section we introduce a minimal Hamiltonian that is sufficient to model the $M_{4,5}$ resonant absorption edge of the Sm ions and fit the model to the experimental absorption represented by TEY spectra.

 Hybridization between the $f$-subsystem and the rest of the conduction electrons in SmB$_6$ is an important interaction responsible for the development of the coherent Kondo state at low temperatures.\cite{Zhang011011, Frantzeskakis2013} The effect of this hybridization is estimated to be of the order $V_{kf}/U_{df}$,\cite{Doniach1982} where the coupling between the valence and f-electrons, $V_{kf}\sim 1$eV, is much smaller than the Coulomb interaction $U_{df}\sim 10$eV. Therefore, we neglect the hybridization and use  a single ion model that includes 14 $f$ and 10 $d$ orbitals, which also would be the minimal orbital set to describe $3d\rightarrow4f$ resonant excitations. 
 
 Considering  the three major interactions---Coulomb con\-tri\-bution, spin--orbit coupling and crystal field effects---the Hamiltonian can be written as\cite{ballhausen1962ligand}
 \begin{equation}
\hat{H} = \hat{H}_\text{Coul} +  \hat{H}_\text{SO} + \hat{H}_\text{CF}.
\end{equation}
The Coulomb part
 \begin{equation}
\hat{H}_\text{Coul} = 
\frac{1}{2}
\sum\limits_{m m' m'' m'''}  U_{m m' m'' m'''}
\hat{c}^\dag_m
\hat{c}^\dag_{m'}
\hat{c}_{m''}
\hat{c}_{m'''},
 \end{equation} 
 responsible for the electronic correlations and the resulting multiplet structure,  is the most significant contributor. In our model, we fully account for the $f$-$f$ and $f$-$d$ interactions. For example, the $f$-$f$ contribution ($l=3$)  has the following form
 \begin{equation}
U_{m m' m'' m'''}^{ff} = 
\sum\limits_{k=0}^{l} a_k(m, m', m'', m''') F_{ff}^{(2k)},
\end{equation}
where all $a_k$ are known coefficients specific to the set of the $f$-orbitals so that $U^{ff}$ is fully determined by the four Slater integrals, $F_{ff}^{(0)}$, $F_{ff}^{(2)}$, $F_{ff}^{(4)}$, $F_{ff}^{(6)}$. The integrals themselves can be estimated from the radial part of the $f$-orbitals\cite{cowan1981theory} and, if necessary, later refined by fitting to experimental data. It is noteworthy that the $F_{ff}^{(0)}$ integral usually cannot be determined reliably due to the screening by the conduction electrons.\cite{Anisimov7570} Since the $F_{ff}^{(0)}$ results in the rigid shift of absorption spectrum as a whole, which will be anyway fitted here, this theoretical challenge does not entail any practical consequence for the current consideration.  In a similar way, the $f$-$d$ part depends on  five other  parameters, controlling so-called direct ($F_{df}^{(2)}$, $F_{df}^{(4)}$) and exchange ($G_{df}^{(1)}$, $G_{df}^{(3)}$, $G_{df}^{(5)}$) Coulomb interactions between $f$- and $d$-orbitals.\cite{ballhausen1962ligand}
 
Within the central field approximation, the spin--orbit interaction  reduces to two constants $\zeta_{3d}$ and $\zeta_{4f}$, determining the strength of the spin--orbit coupling in the $d$- and $f$-shells, respectively:
 \begin{equation}
\hat{H}^\text{SO} = 
\zeta_{3d} \,
\boldsymbol{\hat{L}}_{3d} \cdot \boldsymbol{\hat{S}}_{3d}
+
\zeta_{4f} \,
\boldsymbol{\hat{L}}_{4f} \cdot \boldsymbol{\hat{S}}_{4f}.
\end{equation}

In SmB$_6$, the central Sm ion is surrounded by eight B cages located in the vertices of a cube. This forms the source of a cubic crystal field (CF), which  in the absence of spin--orbit coupling splits the $d$ levels into the $e_g$ doublet and $t_{2g}$ triplet, while the $f$ levels are split into the $a_2$ singlet and $t_1$ and $t_2$ triplets. \cite{Kang2015}  Based on the spatial extent of the radial functions, we can order the total CF splitting of the  3$d$ core and 4$f$ valence levels as $\Delta_\text{cf}(3d) < \Delta_\text{cf}(4f)$. For the homologous compound CeB$_6$, the CF splitting of the 4$f$ levels ($\Gamma_8-\Gamma_7$ splitting of $J=5/2$ $f^1$ configuration) has been determined experimentally to be of the order of 46 meV.\cite{Zirngiebl4052, Loewenhaupt1985245, Alistair066502} Lower values with a  wider spread  ranging from 1 meV to 27 meV were reported for  the  overall CF splitting of the 4$f$ levels in  SmB$_6$.\cite{Nickerson2030, jiao2016, Alekseev1993, Antonov165209} Since the splitting of the energy levels caused by the CF is very small, in fact  comparable to or smaller than the experimental energy resolution of our experimental data (100--200\,meV),  we omit  the CF altogether, similar to other works.\cite{Shick2015} We do so in order to avoid unnecessary numeric overhead, although the effective CF for the 3$d$ and 4$f$ electrons could have be easily included, were it necessary.

The resonant part of the absorption spectrum can thus be calculated up to a proportionality factor 
as 
\begin{multline}
f^\text{res}_{2}(E) 
\iffalse
=
 -\frac{\pi m E^{2}} {\hbar^2}
 \left| \left \langle
R_{3d}(r)| r | R_{4f}(r)
\right \rangle \right|^{2} \\ 
\times
\else
\sim
\fi
\frac{1}{Z}
\sum_{\alpha} e^\frac{-E_\alpha}{k_\text{B}T}
\sum_{\beta} \left| \left \langle
\Psi_\beta(\theta, \varphi)| \vec{\epsilon}\cdot\vec{r} | \Psi_\alpha(\theta, \varphi)
\right \rangle \right|^{2} \\
 \times
 \frac{-1}{\phantom{-}\pi}
\text{Im}\left[\frac{1}{(E_{\beta}-E_{\alpha}-E) + i\frac{\Gamma(E)}{2}}\right].
\label{eq_final}
\end{multline}
Here, the outer sum takes care of the  thermodynamic averaging for a finite temperature $T$, including a possible degeneracy of the initial state $|\Psi_\alpha\rangle$. The inner sum accounts for the transition probabilities between the initial,  $|\Psi_\alpha\rangle$, and the final,  $|\Psi_\beta\rangle$, states, whereas the corresponding  lifetimes are given by the function $\Gamma(E)$, which  in the current consideration is to be determined from fits to experimental data. The polarization of the incident radiation is given by $\vec{\epsilon}$.
 
\begin{figure} 
\includegraphics[width = \columnwidth]{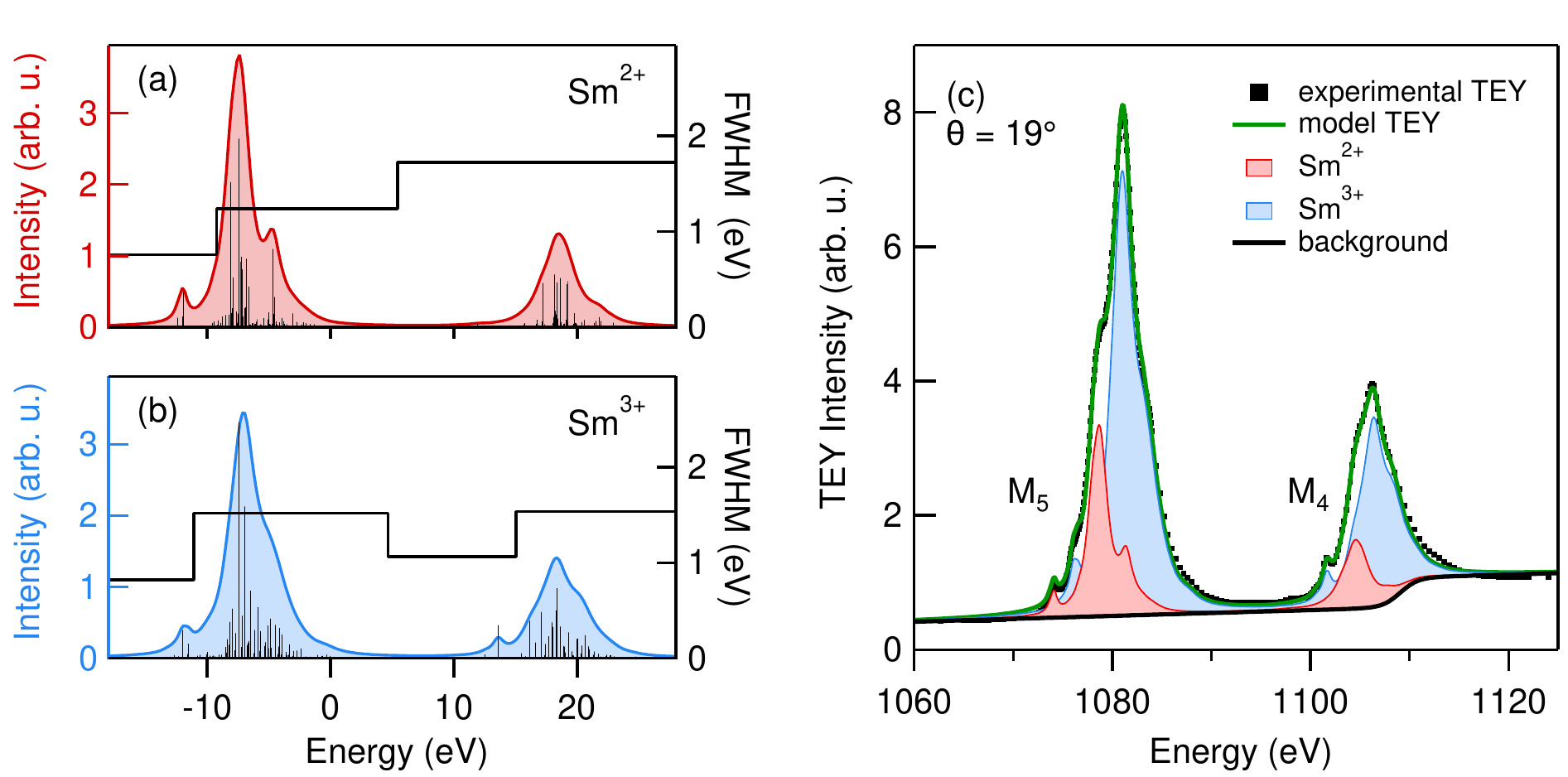}
\colorcaption{(a,b) CFT calculations for Sm$^{2+}$ and Sm$^{3+}$. The spiky spectrum shows the transition energies and their probabilities, as they result from CFT calculation, whereas the continuous curves shows the life-time-broadened spectra. The corresponding energy-dependent life-time, $\Gamma(E)$, is plotted against the right axis. (c) Fit (green line) to the experimental TEY signal (black dots) with the weighted sum of the two Sm components and the smooth background due to the continuum excitations modeled as slope with broadened edges.  Only $M_5$  edge is within the plotted region. }
\label{CFT_figure}
\end{figure}

To perform the actual calculation we used the \textit{Quanty}\cite{Quanty, Haverkort012001, Haverkort165113, Lu085102, Haverkort57004}  framework developed by M. W. Haverkort, which offers a convenient and flexible way to program this quantum mechanical problem in second quantization. The calculated spectra and the optimized input constants are shown in Fig.\,\ref{CFT_figure} and table \ref{params}. As evident from the table,  unlike Ref.\,\onlinecite{Thunstrom2009}, we use two separate sets of Slater $F^{(k)}_{ff}$-integrals and the spin--orbit coupling $\zeta_{4f}$.  The reason for this is the relaxation of atomic orbitals upon the $3d \rightarrow4f$ excitation, accounting for which should improve the match between our  model and the experiment.

Fig.\,\ref{CFT_figure}(a,b) shows the spectra as calculated by the \textit{Quanty} code and broadened according to the energy-dependent life-time, $\Gamma(E)$, plotted against the right axis. The panel (c) demonstrate the actual match  to the experimental TEY data, in which the TEY is modelled as a sum of the 3+ and 2+ components and a smooth background due to the continuum excitations.

As can be seen, the calculated model curves  fit well to  the experimental data, while  the optimized values of the Slater integrals are not  much different  from those reported in an earlier \textit{ab initio} calculation.\cite{Thunstrom2009, Thole5107} Importantly,  we also find that the parameters quoted in Table \ref{params} nicely reproduce  properties of the ground state spectrum, namely the transition energies   $\Delta E({^7F_0} \rightarrow {^7F_1}) = 40$\,meV    in Sm$^{2+}$ and  $\Delta E({^6H_{5/2} } \rightarrow {^6H_{7/2}} ) = 120$\,meV   in Sm$^{3+}$, as measured by inelastic neutron scattering experiments.\cite{Alekseev1993, Alekseev1995, Nefedova1999} Therefore, the constructed model can be considered as a good  approximation to the resonant absorption of Sm$^{2+}$ and Sm$^{3+}$ ions. This is further validated by taking into account that TEY data exhibited little evidence for non-linearity, and as a consequence, for the saturation and self-absorption effects that could  have drawn the TEY signal away from the ``true'' absorption.

The overall angular dependence of the TEY signal is sufficiently involved to merit a separate consideration.\cite{ebel_2004, FRAZER2003161, Schroeder1995L371} To circumvent these difficulties in the next section we use instead x-ray reflectometry, which in addition to chemical and valence profiles allows for extraction of atomic scattering factors in absolute units. 

\begin{figure*}
\includegraphics[width = 0.9\textwidth] {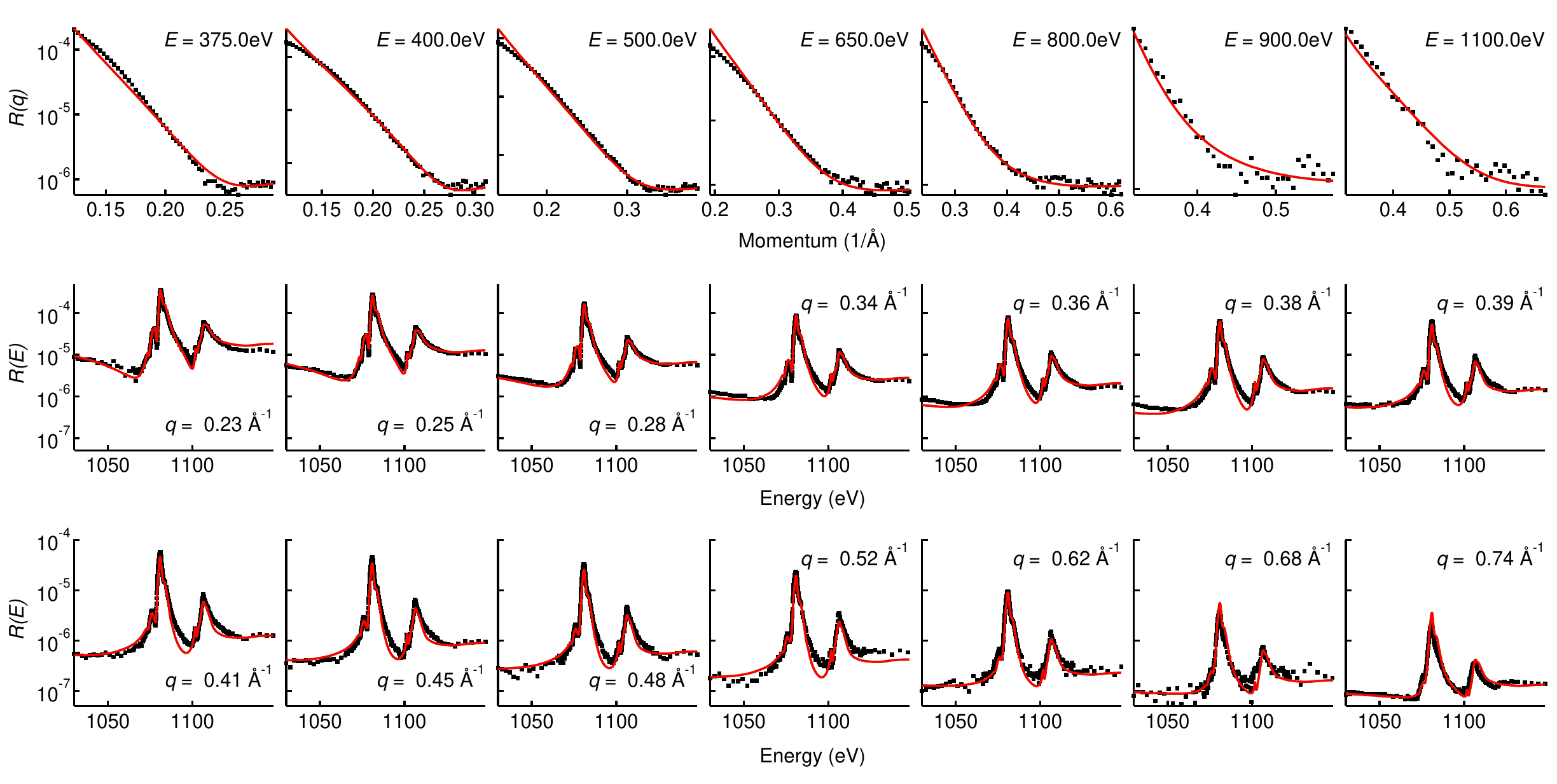}
\colorcaption{ Fits to reflectivity data, measured as off-resonant momentum scans (upper row) or on-resonance (lower two rows) energy scans. The red curves correspond to the calculated reflectivity, while the black symbols show the experimentally measured intensity.  Here, we show representative spectra out of  those used in the actual fit so that all  energies and momentum transfers are uniformly covered. }
\label{reflectivity}
\end{figure*}

\section{Chemical and valence profiling using x-ray reflectometry }
\label{profiling}
To obtain chemical profiles, $c_i(z)$, for  all of the atomic species, $i$, composing the sample and to further refine the on-resonance atomic scattering factors of Sm$^{2+}$ and Sm$^{3+}$ we use resonant x-ray reflectometry,\cite{ZHOU1995223, DIETRICH19951} which is an element and bulk sensitive technique, that previously proved to be useful in determining chemical composition, valence and magnetization profiles, orbital ordering,\cite{Macke14, Nayak13, Nayak15, Filatova185012, Benckiser11} and very recently was used to extract information about the electronic properties of oxide heterostructures.\cite{Borrero2016} 

The method relies on the fact that for a flat layered sample, the intensity of the specularly reflected x-ray beam 
$R(q_z, E)$ 
can be relatively easily obtained from the position- and energy-dependent complex refractive index, $n(z, E)$, which in turn is given by the sum of the contributions of all the atoms composing the material:\cite{Fink056502, Macke14}
\begin{equation}
 n(z, E) =1 -  \frac{r_e \lambda^2 }{2 \pi} \sum\limits_i c_i(z) f_i(E).
\end{equation}
Here $r_e$ is the classical electron radius, $\lambda$ is the wavelength of the exciting radiation, $c_i(z)$ is the spatially varying atomic concentration for the atomic species  $i$ and   $f_i(E)$ is the corresponding total complex atomic scattering factor. The $z$-axis is aligned perpendicularly to the sample surface, while the sample is assumed to be homogeneous in the $x$--$y$ plane.

The generic approach to analyze reflectometry data, $R^\text{exp}(q_z, E)$, is to parametrize both the chemical profiles, $c_i(z)$, and the atomic scattering factors, $f_i(E)$, with models appropriate to the problem at hand and then to optimize all unknown parameters so that the reflectivity, $R^\text{mod}(q_z, E)$, modeled based on  $c_i(z)$ and  $f_i(E)$, fits the experimentally measured spectra $R^\text{exp}(q_z, E)$. To perform this task,  we combine the Parratt formalism \cite{Parratt54} with a differential evolution for optimization of the fit parameters.\cite{Bjorck07, Storn97, Macke14} We parametrize the continuous chemical profiles, $c_i(z)$,  by a set of layers $l=1, 2,... N $ with certain thicknesses $d_{i,l}$, concentrations $c_{i,l}$ and roughnesses $\sigma_{i,l}$.\cite{Macke14} The latter determine how rapidly the concentration changes between neighboring layers.

First, we pin down the parameters driving the  concentration profiles. For this, we use $\theta$--$2\theta$ scans measured at constant energies ranging from $E=375\,\text{eV}$ up to $1200\,\text{eV}$.  At this stage, we rely on the  off-resonant energies, for which we can utilize theoretical scattering factors provided by C. T. Chantler.\cite{Chantler95} Analyzing the experimental data, it turned out that $N=4$ layers were already enough to build a sufficiently detailed representation of $c_i(z)$. The atomic concentrations $c_\text{B,1}$ and $c_\text{Sm,1}$  for the deepest layer $l=1$ were kept fixed at their stoichiometric values calculated from the known bulk crystal structure.\cite{Alekseev0295, Trounov0953} Parameters for the remaining three layers were treated as unknown and were determined from the fits. At this stage, we did not discriminate between Sm$^{2+}$ and Sm$^{3+}$ yet, instead one common atomic concentration $c_{\text{Sm},l} =  c_{\text{Sm}^{2+},l} +  c_{\text{Sm}^{3+},l} $ was used.

In x-ray reflectometry, to distinguish between different atomic species, the atoms have to differ in their scattering factors $f_i(E)$,  due to different oxidation states, different crystal field, or any other reason. That is why away from the resonance energies, Sm$^{2+}$ and Sm$^{3+}$ remain essentially indistinguishable. Therefore, in order to establish separate valence profiles,  $c_{\text{Sm}^{2+},l}$ and  $c_{\text{Sm}^{3+},l}$, one has to move into to the resonance energy range shown in Fig.\,\ref{CFT_figure}. As one may see,   Sm$^{2+}$ and Sm$^{3+}$  slightly differ in position and form of the resonance peaks, which eventually allows separate valence profiles to be extracted from the reflectometry data.  Again, like at the first stage,  we fix the concentrations of the Sm atoms located in the bulk layer, such that the average valency matches the  value of +2.56,  known from literature.\cite{vainshtein1965x, Cohen1970, tarascon1980, Grushko1985, Utsumi155130} For the remaining layers, all their parameters are treated as unknown fit variables.

The total scattering factor in this region is taken as the sum of the off-resonant part, as provided by  C. T. Chantler,\cite{Chantler95} and the resonance part as derived in section\,\ref{CFT} based on CF calculations with the real part of the scattering factor determined via Kramers--Kronig transformation.\cite{Johnson1975} 

In addition to the unknown scaling and energy offset in the resonance part, we also allow for a small energy shift and broadening in the step-like off-resonance part.  

\begin{figure*}
\includegraphics[width =0.99 \textwidth] {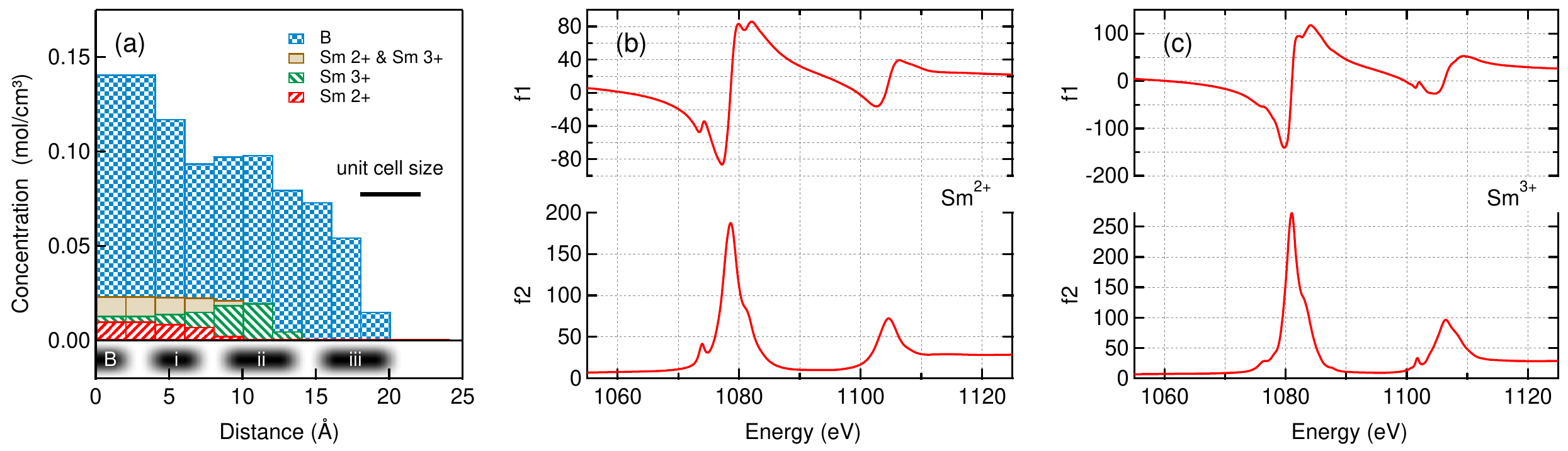}
\colorcaption{(a) The resulting chemical profile, based on which the model reflectivity curves were calculated. (b,c) The atomic scattering factors necessary for calculation of the refractive index were taken from Chantler,\cite{Chantler95} except for the Sm on-resonance region, which derives from the previous section and was further optimized to reflectivity data.}
\label{chemprof}
\end{figure*}

As the last stage, we let all the unknown fit parameters  relax   completely by optimizing the chemical composition profile and  the on-resonance optical constants simultaneously. The resulting fits to reflectivity data are shown in Fig.\,\ref{reflectivity},  whereas the chemical profiles, together with  the atomic scattering factors for Sm$^{2+}$ and  Sm$^{3+}$, are provided in Fig.\,\ref{chemprof}. 

\section{Discussion and comparison to photoemission}
\label{discussion}

Besides the stoichiometric bulk, we can single out three different layers at the surface of the cleaved sample, each approximately one to two unit cells thick. Counting from the bulk (B) on the left (see Fig.\,\ref{chemprof}) these are: (i) A region with a mild prevalence of Sm$^{3+}$ over Sm$^{2+}$ and a slight boron deficiency. (ii) In the second region, the boron deficiency continues to increase, but the major feature here is the almost complete prevalence of Sm$^{3+}$. (iii) The last and outermost region appears to be containing only boron atoms with trace amounts of Sm.  The absence of any substantial quantities of Sm atoms at the very surface is also corroborated by our analysis of the TEY data, as otherwise Sm atoms with broken cubic crystal field should have appeared as additional components in the  SVD decomposition of the surface sensitive TEY data.   

In the following we discuss how the extracted chemical and valence profiles relate to the microscopic structure of the cleaved surface. Had the cleavage occurred primarily between the Sm and B$_6$ layers, \textit{without any considerable loss} of B or Sm atoms,  as reported by the most of the low-temperature STM studies, then the observation of a B-rich surface would seem quite strange. In contrast to these low temperature data, our finding of a boron-rich surface appears to be compatible  with the high-temperature STM,\cite{Ruan136401} where the most probable cleavage plane was concluded to cut through the B$_6$ octahedra so that both counterparts would be B-terminated upon crystal fracture. A preferably B-terminated surface was also reported for low-temperature cleaved SmB$_6$ in a photoemission study,\cite{Heming2014} but the non-stoichiometry at the surface was found to be so profound that possible loss of Sm had to be conjectured. Both loss of Sm and a boron-rich surface were also reported for an electron-beam annealed SmB$_6$ (001) surface,\cite{AONO1979631} suggesting some level of uniformity in experimental observations.  However,  in the most resent photoemission study\cite{Lutz2016} it was possible to select differently terminated  surface spots using an x-ray beam of a few hundred microns in diameter,   which implies that microscopic lateral  inhomogenieties may play an important  role at a cleaved surface of SmB$_6$.   
  
\begin{figure*}
\includegraphics[width =0.99 \textwidth] {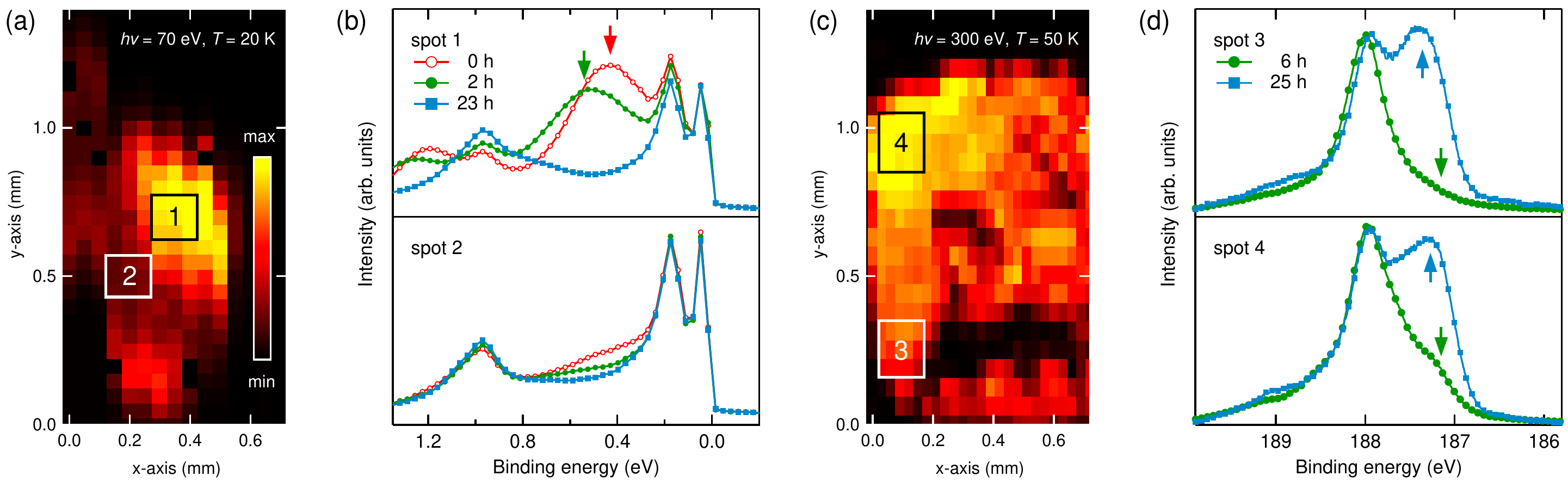}
\colorcaption{ (a) Variation of the photoemission intensity over the cleaved surface, measured at the binding energy of the Sm 4$f$ surface peak ($E_\text{Bind} \sim 0.5$ eV). Based on this intensity map, one can define regions with initially high (spot 1, black square) and low (spot 2, white square) Sm surface signal.  (b) Time evolution of the Sm 4$f$ surface peak averaged over the spot 1 and spot 2. (c)  Intensity map at the B 1$s$ surface peak ($E_\text{Bind} \sim 187.3$ eV).  (d) Corresponding time evolution for the regions with initially low B 1$s$ surface contribution (spot 3) and high contribution (spot 4). The energy position of the respective surface peaks is denoted by the vertical arrows. As time passes on, the inhomogeneity vanishes, both in terms of Sm 4$f$  and B 1$s$ surface contributions.}
\label{ARPES}
\end{figure*}

Unlike  STM, both photoemission and  x-ray reflectivity  are  laterally averaging techniques on a similar length scale of $\sim 100 \mu$m. Considering this similarity, it is insightful  to make a closer comparison of the two experimental methods. Here we perform such a comparison to Al-flux grown SmB$_6$ single crystals cleaved and held in UHV conditions at $T\leq 50$ K. It is notable that right after the cleavage,  one can clearly distinguish between Sm- and B-terminated regions in photoemission signal, though on the time scale of 6--24 h the difference gradually fades away and the surface becomes homogeneous with the boron B 1$s$ states as the only significant surface-related feature. In Fig. \ref{ARPES} we illustrate this evolution  by tracing  
the Sm 4$f$ valence [Fig. \ref{ARPES} (a,b)] and B 1$s$ core [Fig. \ref{ARPES} (c,d)] levels measured with soft x-rays. Besides the sharp bulk peaks, broad surface peaks appear as shoulders at a lower binding energy for B 1$s$ and at a higher binding energy for Sm 4$f$ states. These surface peaks allow us to tag different surface regions as being nominally boron- or samarium-terminated.\cite{Jonathan017038}  The surface peaks exhibit a clear dependence on time,  in general reducing their intensity with the time passed after the cleavage. More precisely, at the Sm-terminated area, we observe a reduction of the Sm surface peak, accompanied by a gradual development of the surface boron peak. The boron-terminated area shows a comparable  progressive development in the surface boron peak, while the Sm surface feature is almost absent. After 23 hours aging at low temperature the spectra from  the two initially different regions become very similar: Boron spectra develop a surface related shoulder at the B 1$s$ peak, but Sm 4$f$ spectra show no significant traces of  the  surface peak [Fig. \ref{ARPES} (b, d)]. 

\begin{figure}[b]
\includegraphics[width =0.8 \columnwidth] {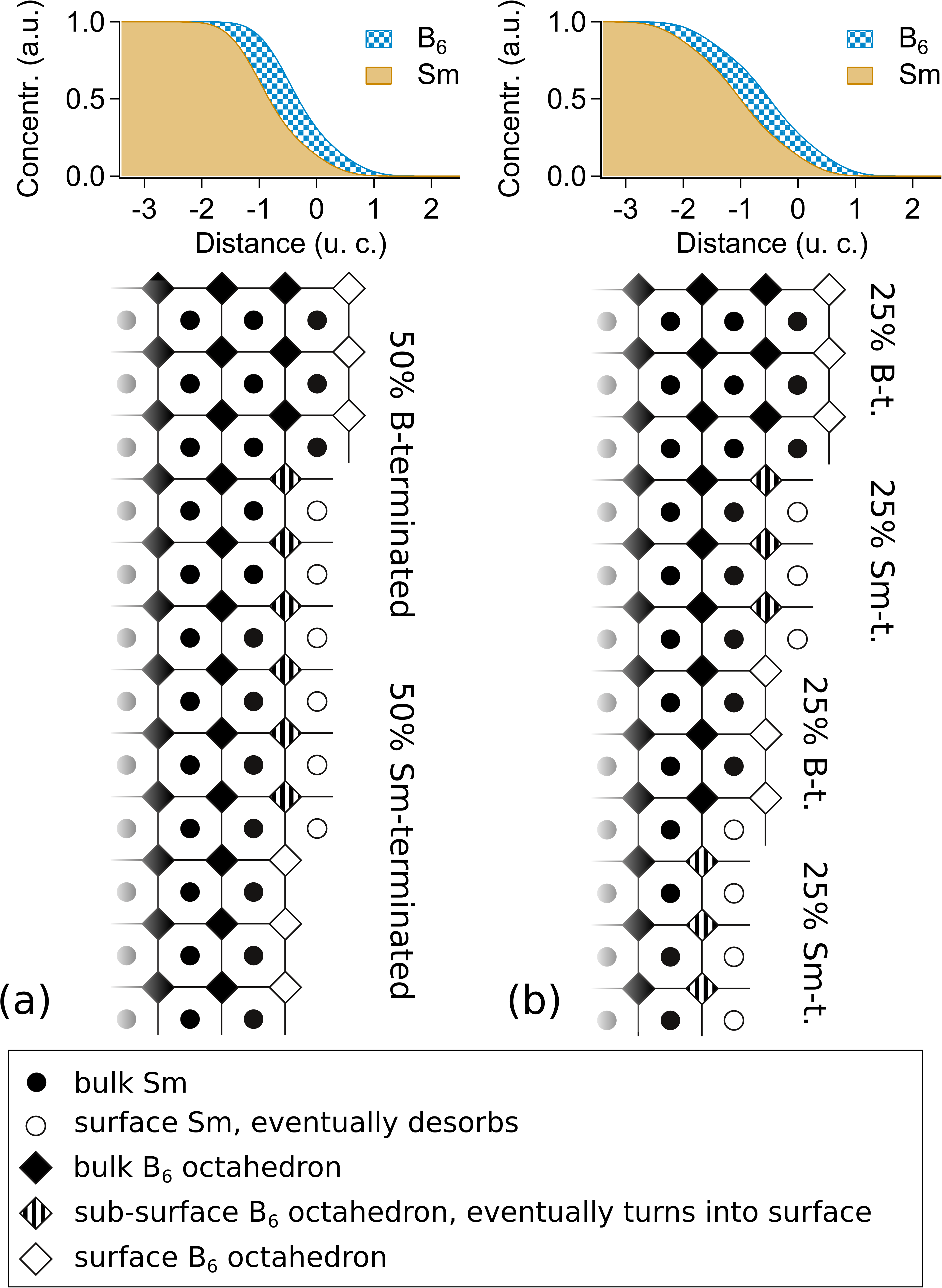}
\colorcaption{Schematic representation of a cleaved surface of SmB$_6$ and the resulting laterally averaged chemical profiles that are attained upon loss of surface Sm atoms. (a) The simplest case. (b) Effect of  surface steppiness or roughness.  }
\label{SmLoss}
\end{figure}

In its outline, this spectral  development can be well understood  in terms of Sm loss, as originally conjectured by Aono \textit{et al.}\cite{AONO1979631} In Fig.\,\ref{SmLoss} we \textit{schematically} illustrate the process and the resulting chemical profiles. It is equally probable to obtain either Sm- or B-terminated patches upon the sample cleavage.  Therefore, if probed selectively in the photoemission experiment, the Sm-terminated patches will initially exhibit a notable  surface-related Sm peak, while the B-terminated patches---a boron surface peak. As time passes on, the surface Sm atoms are gradually lost and the Sm surface peak vanishes. The lack of these atoms will disrupt the underlying boron layer, which in turn leads to the growth of the B surface peak at the former Sm patches.  At the B-terminated patches, in contrast to the Sm patches, no such massive changes are expected. The reason for this lies in the relative stability of the boron sub-lattice and its  protection of the topmost Sm layer so that this surface layer appears indistinguishable from the underlying Sm bulk layers.

Likewise, the  loss of the Sm surface atoms is essential to understanding of the B profile, as extracted from the x-ray reflectometry data.  Immediately upon cleavage, neither Sm nor B should protrude in the laterally averaged chemical profiles. But later,  owing to the  loss of Sm, an ``aged'' surface of SmB$_6$  will eventually exhibit a  B profile that extends beyond that of Sm. A schematic illustration for this can be  found  in Fig.\,\ref{SmLoss} (a).  The schematics can be further improved to cover the less abrupt onset of boron, as seen in the experimental data [Fig.\ref{chemprof} (a)]. For that, one has to  account for the stepped character of the cleaved surface and a possible loss of some B atoms by the damaged surface B$_{6-x}$ octahedra, as shown in Fig.\,\ref{SmLoss}\,(b).  

There is another important comparison to  the photoemission data to be made.  Namely, one has to clarify the apparent disparity of the time scales on which the surface reconstruction occurs. As shown in section \ref{stability},  in the case of the room temperature reflectomentry, the process must have been over by the end of  the sample alignment, which  provides the upper limit of 2 hours  for the reconstruction. On the other hand, in the case of the low temperature photoemission the reconstruction takes  6--24 hours.

Though a deposition of  residual gases  could have been a big problem for the reflectometry analysis, it  is unlikely to play an important role in  the reconstruction, since an accidental deterioration of vacuum does not seem to have any effect on  the pace of the reconstruction.\cite{Lutz2016} As we see from our data, the process appears to be thermally activated, similarly to time-dependent Yb valence drift, observed in another photoemission study.\cite{Reinert12808} This key role of the temperature in the surface reconstruction offers a natural explanation to the differences between the low\cite{Rossler2014, rossler2016surface,jiao2016} and high\cite{Ruan136401} temperature STM measurements.

\section{Summary}
\label{summary}
We performed soft x-ray absorption and reflectometry measurements on SmB$_6$ samples cleaved at room temperature. Having ensured the stability of the cleaved surface we analyzed the absorption data at the  $M_{4,5}$ edge of Sm and showed that there are essentially only two types of Sm ions:  Sm$^{2+}$ and  Sm$^{3+}$, which also suggested that there should be no Sm atoms at the very surface, as these would have different coordination. We used momentum and energy-dependent reflectivity data to extract depth- and element- resolved chemical profiles for both B and Sm, which confirmed a boron domination at the laterally averaged surface. Knowing that there are only two types of Sm ions,  separate profiles were extracted for each valency. To this end the reflectometry data were  backed up with a crystal field calculation, used to model optical properties of the two different Sm ions at the  $M_{4,5}$  resonance. Ultimately, three spatial regions were identified before the bulk properties  recover: a boron-rich topmost layer; an underlying boron-deficient Sm$^{3+}$ layer; and an layer with a mild Sm$^{3+}$ prevalence and slight boron deficiency.  Irrespective of the initial termination,  a boron termination is established eventually,  though the time required may vary substantially. While at room temperature it takes less than two hours, below 50 K the reconstruction occurs on a time scale of about 24 hours.  Thus we conclude that a thermally activated process involving loss of surface Sm should stand behind the observed surface reconstruction. 

We believe that the established chemical and valence profiles will be helpful for a  better understanding and control of the surface polarity \cite{Sawatzky2013} of SmB$_6$ and consequently in interpreting the emergence of surface states in this system.\cite{Kim075131, Trenary2012}

\section{Acknowledgments}
 V.B.Z. and K.F. would like to thank M. W. Haverkort for the insightful discussion and the introduction to CFT. C.H.M thanks Diamond Light Source  and especially T.-L. Lee and C. Schlueter for experimental support at beamline I09 (SI14732).  Likewise C.H.M  thanks J. D. Denlinger and the beamline staff of MERLIN beamline 4.0.3 at Advanced Light Source, which is a DOE Office of Science User Facility under contract no. DE-AC02-05CH11231. 
B.K.C. and B.Y.K. were supported by National Research Foundation of Korea (NRF) grants funded by the Ministry of Education (NRF-2017R1A2B2008538) and Bank for Quantum Electronic Materials-BQEM000652. This work has been supported by the German Research Foundation DFG  through SFB 1170 ``ToCoTronics'' (project C04 \& C06), SFB 1143 ``Correlated magnetism: From Frustration to Topology'' (project C03), and the Natural Sciences and Engineering Research Council of Canada, and the Max Planck-UBC Centre for Quantum Materials. Some experiments for this work were performed at the Canadian Light Source, which is funded by the Canada Foundation for Innovation, NSERC, the National Research Council of Canada, the Canadian Institutes of Health Research, the Government of Saskatchewan, Western Economic Diversification Canada, and the University of Saskatchewan.

%

\end{document}